\documentclass[aps,prb,preprint]{revtex4}
\usepackage{psfrag,graphicx}
\begin{document}

\newcommand{\bra}[1]{\langle #1 |}
\newcommand{\ket}[1]{| #1 \rangle}
\newcommand{\bk}[2]{\bra{#1} #2 \rangle}
\newcommand{\bok}[3]{\bra{#1} #2 \ket{#3}}
\newcommand{\kb}[2]{\ket{#1} \bra{#2}}
\newcommand{\kk}[1]{\kb{#1}{#1}}
\newcommand{\norm}[1]{\{#1\}}
\newcommand{\Half}{\frac{1}{2}}
\title{Calculation of single-beam two-photon absorption rate of lanthanides:
 effective operator method and perturbative expansion}

\author{Chang-Kui Duan}
\affiliation{
Institute of Applied Physics and College of Electronic Engineering,
Chongqing University of Posts and Telecommunications, Chongqing 400065, China.
}
\affiliation{
Department of Physics and Astronomy, University of Canterbury, Christchurch,
New Zealand}
\author{Gang Ruan}
\affiliation{
Institute of Applied Physics and College of Electronic Engineering,
Chongqing University of Posts and Telecommunications, Chongqing 400065, China.
}
\author{Michael F. Reid}
\affiliation{
Department of Physics and Astronomy and MacDiarmid Institute of
Advanced Materials and Nanotechnology, University of Canterbury, Christchurch,
  New Zealand}

\date{\today}

%\vskip 1cm

\begin{abstract}
  Perturbative contributions to single-beam two-photon transition
  rates may be divided into two types. The first, involving low-energy
  intermediate states, require a high-order perturbation treatment, or
  an exact diagonalization. The other, involving high energy
  intermediate states, only require a low-order perturbation
  treatment. We show how to partition the effective transition
  operator into two terms, corresponding to these two types, in such a
  way that a many-body perturbation expansion may be generated that
  obeys the linked cluster theorem and has a simple diagrammatic
  representation.
 
% \pacs{71.20.Eh, 32.70.Cs, 71.27.+a, 32.80.Wr, 04.25.Wx}
\end{abstract}

\maketitle

\section{Introduction}

Two-photon laser spectroscopy is an important complementary technique
to linear spectroscopy because it has a different parity selection
rule, allows access to higher energy states, and has a greater variety
of possible polarization choices than linear
spectroscopy.\cite{Dow1989} The calculation of two-photon absorption
for strongly correlated many-electron systems, especially for rare
earth ions in solids, dated back to Axe's work in 1964.\cite{Axe1964}
However, few quantitative measurements of two-photon absorption were
made until the early 1980's.\cite{DagD1981} Extensive measurements
have been carried out
thereafter.\cite{DowB1982,DowB1983,DowC1983,ChaP1986,GacM1989,Den1991,SytP1993,MurN1997}
Most of these measurements cannot be explained by Axe's lowest-order
calculations, and as a result, many calculations, some using
perturbation theory
\cite{JudP1982,DowB1983,Lea1987,BurR1993,CeuV1996,Sme1998,BryR1998}
and some using full calculations in truncated
spaces,\cite{BurK1993,BurB2003} have been carried out to interpret the
experimental results.  Recently, we have been able to explain the
puzzling two-photon absorption intensities and polarization
dependences of Gd$^{3+}$:LaF$_3$ and Eu$^{2+}$:CaF$_2$ by full
calculations in a truncated $4f^N + 4f^{N-1}5d$ 
space.\cite{DuaR2002,BurB2003}  However, these calculations cannot explain
the two-photon absorption intensities of Sm$^{3+}$, Eu$^{3+}$ and
Tb$^{3+}$ doped in elpasolites.\cite{ThoK1999a,ThoK1999b,MccM2003} It
appears that contributions from high-order perturbations or high
energy intermediate states must be considered.\cite{WanN2003}

A systematic way to calculate properties of many-body systems is using
effective Hamiltonians and operators. These techniques have been
extensively developed in the
literature.\cite{Bra1967,Lin1984,KuoO1990,HurF1993,DuaR2001,KilJ2003a}
The basic idea is to transform the exact time-independent (usually
many-body) Hamiltonian $H$ into an effective Hamiltonian $H_{\rm eff}$
acting on a restricted model space of manageable dimension. The exact
eigenvalues and model space eigenvectors (not the exact eigenvectors)
can be obtained by diagonalizing $H_{\rm eff}$.  For a
time-independent operator $O$, such as a transition moment operator,
an effective operator $O_{\rm eff}$ may be introduced that gives the
same matrix elements between the model space eigenvectors of $H_{\rm
  eff}$ as those of the original operator $O$ between the
corresponding true eigenvectors of $H$.\cite{DuaR2001} Although the
forms of $H_{\rm eff}$ and $O_{\rm eff}$ are generally more
complicated than, respectively, $H$ and $O$, the calculations based on
$H_{\rm eff}$ and $O_{\rm eff}$ have many advantages over variational
and other direct calculations based on $H$ and $O$, such as smaller
bases, less calculation effort, order by order approximations, and the
calculation of all eigenvalues and transition matrix elements
simultaneously for a multi-dimensional model space.  More details on
effective operators can be found in a recent paper of Duan and Reid
\cite{DuaR2001} and references therein.

Many-body perturbation theory (MBPT) may be used to expand effective
Hamiltonians and operators order by order. The calculations are often
represented by Goldstone diagrams.\cite{Gol1957,LinM1985}  The linked
cluster theorem \cite{Bra1967,LinM1985,Lin1974} implies that
disconnected diagrams cancel for effective Hamiltonians and effective
operators, provided that the model space has been carefully chosen and
the model states have been properly orthogonalized.  This cancellation
reduces the number of high order diagrams greatly and ensures size
consistency.  The linked cluster theorem also holds for one-photon
transition operators.\cite{Bra1967,DuaR2001} However, the application
of MBPT to two-photon transitions is more difficult than the case of
one-photon transitions, in that there are energy denominators in
two-photon transition operators that contain both the photon energies
and exact electronic energies.\cite{BryR1998,DuaR2002b} Also, the
intermediate states can be any eigenstates of the system, including
states in the model spaces containing initial and final states and
other low excited states, making the energy denominators change
drastically and makes perturbative expansions for these intermediate
states impossible.

In this paper, we explore the effective operator method for two-photon
transition calculations by combining exact calculations in a truncated
space \cite{BurK1993,BurB2003} with perturbative methods for the rest
of the states.\cite{BryR1998,DuaR2002b} In section II we review the
basic formalism for effective operator methods; In section III the
partition of two-photon transition operator is given in detail;
Section IV presents the perturbation expansion that may be suitable
for diagram representation and applying linked cluster theorem.  The
diagram representation and diagram evaluation rules and linked cluster
theorem themselves for effective two-photon transition operator are
highly nontrivial and will be presented in a followed paper.

\section{Basic formalism}

Most of the formalism required has been treated in detail in a
monograph by Lindgren and Morrison\cite{LinM1985} and more recently
summarized by Killingbeck and Jolicard.\cite{KilJ2003a} The concept
of biorthogonal model space eigenvectors has been summarized by Duan
and Reid.\cite{DuaR2001} Here shall only give a brief description of
the formalism necessarily in the presentation that follows.

The time-independent Hamiltonian $H$ is written
as the sum of a model Hamiltonian $H_0$ and a perturbation V
\begin{equation}
H = H_0 + V.
\end{equation}
Usually $H_0$ is chosen in such a way that its eigenvalues and
eigenvectors can be obtained more easily than for $H$. For example,
when $H$ is the Hamiltonian for a many-body system, $H_0$ is
usually chosen such that each particle moves independently in the
average field of other particles and $V$ denotes the remainder of $H$.

A complete set of orthonormal eigenvectors $\{\ket{\alpha}\}$ and
corresponding eigenvalues $\{E^{\alpha}_0\}$ for $H_0$ are assumed to
be available
\begin{eqnarray}
H_0  \ket{\alpha} &=& E^{\alpha}_0\ket{\alpha},\\
\bk{\alpha}{\beta} &=&\delta_{\alpha \beta}.
\end{eqnarray}
A general model space $P_0$, often referred to as a quasi-degenerate model
space, is defined as the space spanned by $d$ successive eigenvectors
of $H_0$ (not necessarily strictly degenerate). The remaining part of the Hilbert
space is called the orthogonal space $Q_0$. Here we denote the associated  
projection operators also as $P_0$ and $Q_0$:
\begin{eqnarray}
P_0 &=& \sum\limits_{\alpha =1}^d \kk{\alpha},\\
Q_0 &=& \sum\limits_{\alpha>d}\kk{\alpha} = 1 - P_0
\end{eqnarray}
It has been shown that $d$ eigenvectors of the full Hamiltonian
$\ket{\Psi^{\alpha}}$ ($\alpha = 1, \cdots, d$) can usually be
projected into the model space as $d$ linearly independent functions
$\ket{\Psi^{\alpha}_0}$ in $P_0$.\cite{KuoO1990,SchW1973} The wave
operator $\Omega_P$ is defined as an operator that transforms all the
$d$ model functions back into the corresponding exact eigenvectors.
\begin{eqnarray}
\ket{\Psi^{\alpha}_0} = P_0 \ket{\Psi^{\alpha}}
\label{model_func}\\
\ket{\Psi^{\alpha}} = \Omega_P \ket{\Psi^{\alpha}_0}.
\label{Omega}
\end{eqnarray}
Note that $\ket{\Psi^{\alpha}_0}$s are not necessarily eigenstates of
$H_0$, but a linear combination of those eigenstates of $H_0$ in
$P_0$. We shall call the space spanned by the $d$ exact eigenvectors
$P$.  The wave operator $\Omega_P$ satisfied the ordinary Bloch
equation\cite{Blo1958}
\begin{equation}
\label{bloch}
[\Omega_P, H_0] = (V\Omega_P - \Omega_P P_0 V \Omega_P )P_0.
\end{equation}

Usually the $\ket{\Psi^{\alpha}_0}$ are not orthogonal but are chosen
to be normalized to unity.  As a consequence $\ket{\Psi^{\alpha}}$ is
not normalized to unity. Denote
\begin{equation}
\bk{\Psi^{\alpha}}{\Psi^{\beta}} = N _{\alpha}^2 \delta _{\alpha   \beta} .
\label{Nalpha}
\end{equation}

An effective Hamiltonian $H_{\rm eff}$ acting on the model space,
which gives the $d$ exact eigenvalues and model eigenvectors upon
diagonalizing, can now be defined. Its form and eigen-equation are
\begin{eqnarray}
&&H_{\rm eff} = P_0H\Omega_P P_0 = P_0H_0P_0 + P_0 V \Omega_P P_0\label{Ham_eff},
\\
&&H_{\rm eff} \ket{\Psi^{\alpha}_0}_k = E^{\alpha}\ket{\Psi^{\alpha}_0}_k.
\label{Sch_eff}
\end{eqnarray}
Instead of calculating $\Omega_P$ directly from (\ref{bloch}), the
effective Hamiltonian are usually calculated via perturbation theory
or phenomenological method and then diagonalized to give eigenvalues
$E^{\alpha}$ and eigenvectors $\ket{\Psi^{\alpha}_0}_k$, It is
straightforward to find from a set of vectors
$_b\bra{\Psi^{\alpha}_0}$ in the model space such that
\begin{eqnarray}
_b\bk{\Psi^{\alpha}_0}{\Psi^{\beta}_0}_k = \delta _{\alpha \beta}
\\
\ket{\Psi^{\alpha}_0}_k~ _b\bra{\Psi^{\beta}_0} = P_0
\end{eqnarray}
It is straightforward to show that
\begin{eqnarray}
_b \bra{\Psi_0^{\alpha}}(\Omega_P^+\Omega_P)^{-1} \in P_0,\\
_b \bok{\Psi^{\alpha}_0}{(\Omega_P^+\Omega_P)^{-1}
   \Omega_P^+\Omega_P}{\Psi^{\beta}_0}_k = \delta _{\alpha \beta},\\
_b\bra{\Psi^{\alpha}_0}H_{\rm eff} = E^{\alpha} ~_b\bra{\Psi^{\alpha}_0},
\end{eqnarray}
which together with Eq. \ref{Omega} and Eq. \ref{Nalpha} show that
\begin{eqnarray}
_b \bra{\Psi^{\alpha}_0} (\Omega_P^+\Omega_P)^{-1}\Omega_P^+ 
    = N^{-2}_{\alpha} \bra{\Psi^{\alpha}_0},\\
P=\sum\limits_{\alpha}\Omega_P P_0 (\Omega_P^+\Omega_P)^{-1}\Omega^{+}_P.
\end{eqnarray}
The transition matrix element of an operator $O$ between states 
$\bra{\Psi^{\alpha}}$ ($\alpha \in A$) and $\bra{\Psi^{\beta}}$ ($\beta \in B$)
is
\begin{equation}
O_{\alpha \beta} = \frac{\bok{\Psi^{\alpha}}{O}{\Psi^{\beta}}}{N_{\alpha}N_{\beta}}.
\end{equation}
Defining
\begin{equation}
O_{\rm eff} = (\Omega_P^+\Omega_P)^{-1}\Omega_P^+O\Omega_P,
\end{equation}
it can be shown that
\begin{eqnarray}
&&_b\bok{\Psi^{\alpha}_0}{O_{\rm eff}}{\Psi^{\beta}_0}_k 
   = \frac{N_{\beta}}{N_{\alpha}} O_{\alpha\beta},\\
&&O_{\alpha\beta} = (_b\bok{\Psi^{\alpha}_0}{O_{\rm eff}}{\Psi^{\beta}_0}_k 
                     ~_b\bok{\Psi^{\beta}_0}{O_{\rm eff}}{\Psi^{\alpha}_0}_k^*)^{-1/2}.
\label{Oeff}
\end{eqnarray}

\section{Partition of effective transition operators for single
beam two-photon absorption}

The perturbative expansion of a general two-beam two-photon transition
rate has been discussed by Duan and Reid\cite{DuaR2002b}. Here we
develop the perturbative expansion specialized for the single-beam
case. We will see that the calculation can be greatly simplified.

The line strength is proportional to the square modulus of following expression
\begin{eqnarray}
T_{fi} &=& \frac{\bok f O k \bok k O i}{E_i+\omega -E_k} 
\label{TPoper0}\\
      &=& {\bok f {O \frac{1}{(E_f+E_i)/2-H}O} i}.
\label{TPoper}
\end{eqnarray}
where $\ket{f}~(f\in F)$, $\ket{i}~(i\in I)$ and 
$\ket{k}(k\in F\cup I\cup K)$ are normalized
exact eigenstates of the systems, $E_f,~E_i,~E_k$
 are the corresponding exact eigenvalues,
$F$, $I$ are the set of final and initial states 
of the transition. $K$ is the set of states of the 
system not included in $I$ and $F$. 
The intermediate eigenstates $k$
 can be any eigenstates of the system, including eigenstates in the sets
 of initial and final states. 
The calculation of $T_{fi}$ can be divided into two
 terms, a term $T_1$ with ``small'' denominators, where
$k\in I\cup F$, and a term $T_2$ with ``large'' denominators, 
where $k\in K$. The operator can be formally written  as
\begin{eqnarray}
T&=& T^1 + T^2\\
\label{term1}
T^1 &=& \sum\limits_{f,i}\kb{f}{i}\sum\limits_{k\in I \cup F}
        \frac{O_{fk}O_{ki}}{(E_f+E_i)/2-E_k}\\
\label{term1b}
    &=& O\frac{P_{I\cup F}}{(H^F+H^I)/2-H}O\\
\label{term2}
T^2 &=&  \sum\limits_{f,i}\kb{f}{i}\sum\limits_{k\in I \cup F} 
         \frac{O_{fk}O_{ki}}{(E_f+E_i)/2-E_k}\\
\label{term2b}
    &=& O\frac{P_K}{(H^I+H^F)/2-H}O.
\end{eqnarray}
Note that $H^F$ and $H^I$ are actually acting on the transition 
final (on the leftmost) and initial states 
(on the rightmost)  and the equalities in 
(\ref{term1b}) and (\ref{term2b})  above are for notational
convenience.

The effective operator for $T^1$ is 
\begin{widetext}
\begin{eqnarray}
&&T^1_{\rm eff} = \sum\limits_{f,i}\ket{\Psi^f_0}_k~_b\bra{\Psi^i_0}
 \sum\limits_{m\in F\cup I} \frac{_b\bok{\Psi^f_0}{O_{\rm eff}}
  {\Psi^m_0}_k~_b\bok{\Psi^m_0}
     {O_{\rm eff}}{\Psi^i_0}_k}{(E_f+E_i)/2-E_m } . 
\end{eqnarray}
\end{widetext}
Note that in this expression we use the exact eigenvalues and model
space eigenvectors.  The matrix elements of the effective operators
are between states in spaces $P_{F0}$ and $P_{I0}$ (which may be the
same).  The matrix elements may be calculated from the effective
Hamiltonian $H_{\rm eff}$ and effective operator $O_{\rm eff}$. 

The effective operator for $T_2$ can be formally written as
\begin{widetext}
\begin{equation}
T^2_{\rm eff} = (\Omega^+_F\Omega_F)^{-1}\Omega_F^+O\Omega_KP_K^0
                \frac{1}{(H^F_{\rm eff}+H^I_{\rm eff})/2-H^K_{\rm eff}}P_K^0 
                (\Omega_K^+\Omega_K)^{-1}\Omega_K^+O\Omega_I.
\end{equation}
\end{widetext}
Once again, $H^F_{\rm eff}$ and $H^I_{\rm eff}$ act on the bra
and ket model space respectively.

The space $K$ includes all other states of the system than those
limited number of states in $I$ and $F$. It is usually of infinite
dimension and the calculation of $H_{\rm eff}^K$, $O_{\rm eff}(F,K)$
and $O_{\rm eff}(K,I)$ is usually impractical or at least very
tedious.  Perturbative expansions giving in the following section can
be used to calculate $T^2_{\rm eff}$ by an order-by-order
approximation.

\section{Perturbative expansion}

Perturbative expansions of $H_{\rm eff}$ and $O_{\rm eff}$ have been
discussed in, for example, Refs. \onlinecite{LinM1985} and
\onlinecite{DuaR2001} by applying the Bloch equation iteratively. Here
we expand $T^2_{\rm eff}$ by perturbation theory to avoid direct calculation of
$H^{\rm eff}$ and $O^{\rm eff}$ in model space $K$, which is usually
of infinite dimension.

Defining 
\begin{eqnarray}
\label{solvent}
S &=& 1/ [(H^F_0 + H^I_0)/2 - H^K_0],\\
\label{deltav}
\Delta V &=& V^K_{\rm eff} - \frac{1}{2}(V^F_{\rm eff} + V^I_{\rm eff}),
\end{eqnarray}
where 
\begin{eqnarray}
V^{\xi}_{\rm eff} &=& P_{\xi 0}V\Omega_{\xi}P_{\xi 0}\\
H_0^{\xi} &=& P_{\xi 0} H_0 P_{\xi 0} 
\end{eqnarray}
act on the transition initial, final and intermediate states for 
$\xi = I,~F,~K$ respectively. The energy denominator can be expanded as follows
\begin{eqnarray}
&& \frac{1}{(H^F_{\rm eff}+H^I_{\rm eff})/2-H^K_{\rm eff})}
= S\sum\limits_{n=0}^{\infty} (\Delta V S)^n\\
&&= S + S V^K_{\rm eff} S - \frac{1}{2} V^F_{\rm eff}S^2 -
\frac{1}{2} S^2V^I_{\rm eff} + \cdots
\end{eqnarray}

Using the Bloch equation, $\Omega_K$ can be expanded as follows
\begin{eqnarray}
&&\Omega_K = P_{K_0} + R_K(V\Omega_K - \Omega_KV\Omega_K)\\
&&        =  P_{K_0} + R_KVP_{K0} + R_KVR_KVP_{K0} - R^2VP_{K0}^{\prime}VP_{K0},
\end{eqnarray}
where 
\begin{equation}
R_K = \frac{P_{(I\cup F) 0}}{H^K_0 - H^{I\cup F}_0}. 
\end{equation}
Using the above expressions, the matrix elements of the zeroth
 and first-order of $T^2_{\rm eff}$ 
between eigenstates of $H_0$, $\ket{f}$ ($f \in P_{F0}$) and $\ket{i}$ ($i\in P_{I0}$) are:
\begin{widetext}
\begin{eqnarray}
&&\bok{f} {T^2_{\rm eff,0}} {i} =
   \sum\limits_{k\in P_{K0}}
    \frac {\bok {f}{O}{k}\bok {k}O{i}}
          {(E_{f0} + E_{i0} )/2 - E_{k0}}\\
&&\bok f {T^2_{\rm eff,1}} i =
  \left \{
  \sum\limits_{k\in P_{K0}}\left [\sum\limits_{l \in Q_{F0}}
  \frac{\bok f V l \bok l O k \bok k O i}
       {(E_{f0}-E_{l0})[(E_{f0}+E_{i0})/2-E_{k0}]}
% the following line should be deleted for nonpreprint 
       \right . \right . \nonumber \\  && \left . \left .
+\sum\limits_{l \in Q_{I0}}
  \frac{\bok f O k \bok k O l \bok l V i}
       {(E_{i0}-E_{l0})[(E_{f0}+E_{i0})/2-E_{k0}]} \right ]
       \right . \nonumber\\
&& \left . +\sum\limits_{l \in Q_{K0}}\sum\limits_{k\in P_{K0}} \left [
  \frac{\bok f O l \bok l V k \bok k O i}
       {(E_{k0}-E_{l0})[(E_{f0}+E_{i0})/2-E_{k0}]}
  +\frac{\bok f O k \bok k V l \bok l O i}
       {(E_{k0}-E_{l0})[(E_{f0}+E_{i0})/2-E_{k0}]} \right ]
 \right . \nonumber\\
&&\left .+ \sum\limits_{k_1 \in P_{K0}}\sum\limits_{k_2\in P_{K0}}
  \frac{\bok f O {k_1} \bok {k_1} V {k_2} \bok {k_2} O i}
       {[(E_{f0}+E_{i0})/2-E_{k_10}][(E_{f0}+E_{i0})/2-E_{k_20}]}
\right . \nonumber\\
&&\left . 
-\frac{1}{2} \sum\limits_{f^{\prime} \in P_{F0}}\sum\limits_{k\in P_{K0}}
  \frac{\bok f V {f^{\prime}} \bok {f^{\prime}} O {k} \bok {k} O i}
       {[(E_{f0}+E_{i0})/2-E_{k0}][(E_{f^{\prime}0}+E_{i0})/2-E_{k0}]}
\right . \nonumber\\
&&\left . -\frac{1}{2} \sum\limits_{i^{\prime} \in P_{i0}}\sum\limits_{k\in P_{K0}}
  \frac{ \bok {f} O {k} \bok {k} O {i^{\prime}} \bok{i^{\prime}} V {i}}
       {[(E_{f0}+E_{i0})/2-E_{k0}][(E_{f0}+E_{i^{\prime}0})/2-E_{k0}]}
\right \}
\end{eqnarray} 
\end{widetext} 
where all eigenvectors and energies are for model Hamiltonian $H_0$
and matrix elements are between eigenvectors of $H_0$.  The transition
rates can then be calculated straightforwardly from Eq.\ref{Oeff}
since the model space eigenvectors are assumed to have already been
calculated from Eq.\ref{Sch_eff}.

Terms of second order or higher in $V$ can also be obtained
straightforwardly.  There are about 20 second order terms but around
100 third order terms. Fortunately, with a suitable partition of $T$
into $T^1$ and $T^2$, usually only the zeroth and first order terms of
$T^2$ need to be calculated, except when zeroth and first -order terms
become zero due to selection rules. In such cases the number of
nonzero second order terms is often greatly reduced.

\section{Conclusion}

A method to calculate single-beam two-photon absorption transition
rates for many-electron systems has been developed using effective
operator methods together with many-body perturbation theory. In this
method the contributions to two-photon transition operator are
partitioned into two terms, one with small drastically varying
denominators, which is treated by doing an exact calculation in
truncated spaces and the other with numerous intermediate states and
large energy denominators, which is treated systematically with
many-body perturbation theory.  Compared to previous methods, the
method presented here has the accuracy of full calculation for
contributions due to drastic-varying low-energy intermediate states
and the simplicity of low-order many-body perturbation theory for
contributions due to high energy intermediate states. It is
also expected that there are linked diagram representations for
the order-by-order expansion.

\section*{Acknowledgement}

CKD and GR acknowledge support of this work by the Natural Science
Foundation of China, Grant No. 10274079 (2002).

\bibliography{302441JCP}

\end{document}